\documentclass[twocolumn]{aastex62}

\usepackage{amsmath}
\usepackage{color}
\graphicspath{{./}{figures/}}
\usepackage{natbib}
\usepackage{enumitem}
\usepackage{float}

\shorttitle{The Stellar Dispersion of NGC1052-DF2}
\shortauthors{Danieli et al.}

\begin{document}

\newcommand\XXX[1]{{\textcolor{red}{\textbf{x\ #1\ x}}}}

\gdef\kms{km\,s$^{-1}$}

\title{Still Missing Dark Matter: KCWI High-Resolution Stellar Kinematics of NGC1052-DF2}

\email{shany.danieli@yale.edu, shanyi1@gmail.com}

\author[0000-0002-1841-2252]{Shany Danieli}
\affil{Department of Physics, Yale University, New Haven, CT 06520, USA \\}
\affil{Yale Center for Astronomy and Astrophysics, Yale University, New Haven, CT 06511, USA \\}
\affil{Department of Astronomy, Yale University, New Haven, CT 06511, USA \\}

\author[0000-0002-8282-9888]{Pieter van Dokkum}
\affiliation{Department of Astronomy, Yale University, New Haven, CT 06511, USA \\}

\author[0000-0002-1590-8551]{Charlie Conroy}
\affiliation{Harvard-Smithsonian Center for Astrophysics, 60 Garden Street, Cambridge, MA, USA\\}

\author[0000-0002-4542-921X]{Roberto Abraham}
\affiliation{Department of Astronomy and Astrophysics, University of Toronto, Toronto ON, M5S 3H4, Canada\\}
\affiliation{Dunlap Institute for Astronomy and Astrophysics, University of Toronto, Toronto ON, M5S 3H4, Canada\\}

\author[0000-0003-2473-0369]{Aaron J. Romanowsky}

\affiliation{Department of Physics and Astronomy, San Jos\'e State University, San Jose, CA 95192, USA\\}
\affiliation{University of California Observatories, 1156 High Street,
Santa Cruz, CA 95064, USA\\}
\begin{abstract}

The velocity dispersion of the ultra diffuse galaxy NGC1052-DF2 was found to be $\sigma_{\rm gc}=7.8^{+5.2}_{-2.2}$\,\kms,  much lower than expected from the stellar mass -- halo mass relation and nearly identical to the expected value from the stellar mass alone. This result was based on the radial velocities of ten luminous globular clusters that were assumed to be associated with the galaxy. A more precise measurement is possible from high resolution spectroscopy of the diffuse stellar light. Here we present an integrated spectrum of the diffuse light of NGC1052-DF2 obtained with the Keck Cosmic Web Imager, with an instrumental resolution of $\sigma_{\rm instr}\approx 12$\,\kms. The systemic velocity of the galaxy is $v_{\rm sys}=1805\pm 1.1$\,\kms, in very good agreement with the average velocity of the globular clusters ($\langle v_{\rm gc}\rangle = 1803\pm 2$\,\kms). There is no evidence for rotation within the KCWI field of view. We find a stellar velocity dispersion of $\sigma_{\rm stars}=8.5^{+2.3}_{-3.1}$\,\kms, consistent with the dispersion that was derived from the globular clusters. The implied dynamical mass within the half-light radius $r_{1/2}=2.7$\,kpc is $M_{\rm dyn}= (1.3 \pm 0.8) \times 10^8 $\,M$_{\odot}$, similar to the stellar mass within that radius ($M_{\rm stars}=(1.0 \pm 0.2) \times 10^8 \ \mathrm{M}_{\odot}$). With this confirmation of the low velocity dispersion of NGC1052-DF2, the most urgent question is whether this ``missing dark matter problem'' is unique to this galaxy or applies more widely.
\end{abstract}

\keywords{galaxies: individual (NGC1052-DF2) -- galaxies: kinematics and dynamics}

\section{Introduction} \label{sec:intro}

A remarkable result of the past twenty years is the apparent regularity of galaxy formation as reflected in the existence of a well-defined relation between galaxy and halo mass with small scatter, the stellar mass -- halo mass relation (\citealt{2006MNRAS.368..715M}, \citealt{2010ApJ...710..903M}, \citealt{2010ApJ...717..379B}, \citealt{2018ARA&A..56..435W}).
The scatter in this relation constrains the possible evolutionary histories of galaxies, and has been measured fairly well at high masses
 (\citealt{2004MNRAS.353..189V}, \citealt{2016ApJ...833....2G}).
However, this scatter is relatively unconstrained at low masses, in the regime where galaxy formation is thought to be less efficient.
Measuring or constraining the halo masses of low mass galaxies therefore provides important information on the scatter in the stellar mass -- halo mass relation and on the question whether galaxy formation is less regulated, or even
stochastic, at low masses.

\begin{figure*}[t!]
{\centering
  \includegraphics[width=180mm]{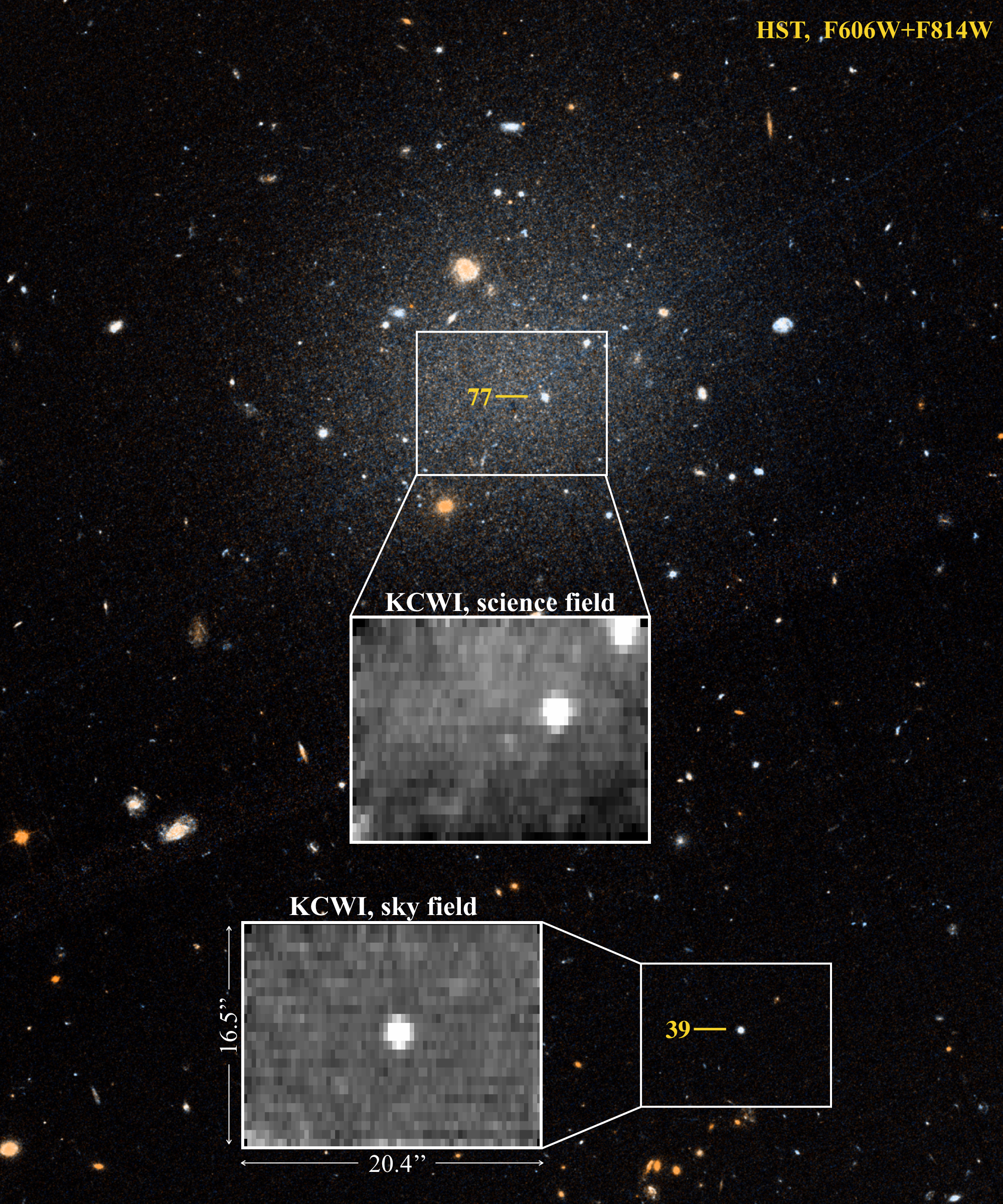}
  \caption{\textit{HST}/ACS color image of NGC1052-DF2, created from the $V_{606}$ and $I_{814}$ bands. The white frames represent the two 
  $20\farcs 4 \times 16\farcs 5$ KCWI pointings.  The upper frame covers the diffuse light of the galaxy out to $0.7R_{\mathrm{eff}}$. The lower frame was used for sky modeling. The insets show the collapsed summed KCWI data cubes.}
  \label{fig:data}
}
\end{figure*}

There is a rich literature on halo mass measurements of low mass galaxies ($\mathrm{M}_{*} \sim 10^8 \ \mathrm{M}_{\odot}$ or lower) in the Local Group (e.g., \citealt{1983ApJ...266L..11A}, \citealt{2009ApJ...692.1464G}, \citealt{2016MNRAS.458L..59M}, \citealt{2018arXiv181104082T}). These studies use the velocities of individual stars to infer the kinematics, the mass density profile, and the halo mass. Other studies focus on gas-rich dwarf galaxies within a few Mpc, inferring the halo mass from H$\alpha$ and/or H\,{\sc i} rotation curves (\citealt{2001AJ....122.2396D}, \citealt{2001MNRAS.325.1017V}, \citealt{2016MNRAS.460.3610O}). There seems to be considerable scatter in the halo mass in this stellar mass regime, although this may partly reflect uncertainties in inclination (see \citealt{2016MNRAS.460.3610O}) and variation in the inner density profiles of halos rather than in total halo masses. 

Outside the Local Group, much less is known about the variation in the stellar mass -- halo mass relation.
The recently identified population of ultra diffuse galaxies (UDGs; \citealt{2015ApJ...798L..45V}) holds the promise of new constraints, as their large spatial extent and often significant globular cluster populations provide probes on  spatial scales
where dark matter should dominate the kinematics.  Using the velocities of globular clusters (\citealt{2016ApJ...819L..20B}, \citealt{2018ApJ...856L..31T}) and stellar velocity dispersions (\citealt{2016ApJ...828L...6V}), UDGs are gradually adding to the sample of low mass galaxies with constraints on their dark matter content beyond the Local Group.

Recently, a relatively nearby UDG at $20 \ \mathrm{Mpc}$, NGC1052-DF2, was inferred to have little or no dark matter, deviating by an unprecedented amount from the expected  $M_{\mathrm{halo}}/M_{\mathrm{stars}}$ ratio. The constraints on the NGC1052-DF2 halo mass were derived by measuring the velocities of ten globular clusters that were assumed to be associated with the galaxy itself. The velocity dispersion of the 10 clusters is $\sigma_{\rm gc}=7.8^{+5.2}_{-2.2}$\,\kms (\citealt{2018RNAAS...2b..54V}). Due to the small number of tracers, the results have a large random uncertainty (see also \citealt{2018ApJ...859L...5M}), may suffer from small sample bias in the likelihood estimator (see \citealt{2018MNRAS.tmp.2765L}), and are sensitive to systematic errors in individual measurements (as demonstrated by the cluster GC-98; see \citealt{2018RNAAS...2b..54V}). 

A more precise way of constraining the kinematics is by measuring the \textit{stellar} velocity dispersion of the galaxy. This is challenging because of the low surface brightness of NGC1052-DF2 and because a relatively high spectral resolution is required. 
The observed broadening of spectral features is $\sigma_{\rm obs}^2 = \sigma_{\rm instr}^2 + \sigma_{\rm stars}^2$; because of this quadratic behavior a spectral resolution $\sigma_{\rm instr} \sim \sigma_{\rm stars} \sim 10$\,\kms \ is required. This is now possible with the Keck Cosmic Web Imager (KCWI), a new instrument on the Keck II telescope that is optimized for precision sky limited spectroscopy of low surface brightness phenomena at relatively high spectral resolution.

\section{Observations and Data Reduction} \label{sec:data}

\subsection{KCWI Spectroscopy}

We obtained IFU spectroscopy of NGC1052-DF2 on 2018 October 11 with the Keck Cosmic Web Imager (KCWI, \citealt{2012SPIE.8446E..13M}, \citealt{2018ApJ...864...93M}) on Keck II. The highest resolution KCWI configuration was chosen where the spectra are still (nearly) sky limited. The medium slicer was used with the BH3 grating, for a field of view of $16.5'' \times 20.4''$. The central wavelength was set to $\lambda_{\rm cen}=5080$\,\AA. The spectral resolution, as measured from arc lamps, ranges
from $14$\,\kms at $\lambda=4800$\,\AA\ ($R\approx 9100$)
to $11$\,\kms at
$\lambda=5300$\,\AA\ ($R\approx 11,600$). 

NGC1052-DF2 is larger than the KCWI field of view, which means that offset exposures have to be used to characterize the sky emission. In practice we alternated between two positions. In the first, ``science" exposures were taken with the KCWI Field Of View (FOV) placed just south-west of the center of NGC1052-DF2, covering the stellar component of the galaxy out to $0.7 R_{\rm eff}$ as well as GC-77, the second-brightest globular cluster associated with the galaxy. 
In the second pointing, ``sky'' exposures were taken with the FOV placed on a field $1\farcm 3$ away, centered on the globular cluster GC-39. The globular cluster takes up only a small fraction of the KCWI area, and is masked in the sky analysis.  The two pointings, along with stacked collapsed images of the science and sky exposures, are shown in Figure \ref{fig:data}.

We obtained exposures of $1,200 \ \mathrm{s}$ at each position, for a total of $9,600 \ \mathrm{s}$ on the galaxy  and $10,800 \ \mathrm{s}$ on the offset field. The total science + sky time that is used in the analysis is 5.6 hrs.
Conditions were somewhat variable, with thin cirrus present during most of the observations.

\subsection{Reduction}
\label{sec:reduction}

The KCWI Data Extraction and Reduction Pipeline (KDERP), with default settings, is used to perform basic reduction and calibration of the data (\citealt{2018ApJ...864...93M}). Each of the 17 science and sky frames is treated independently.  A combination of ``bars'' exposures,
arc lamps, and the science data is used to derive distortion corrections and wavelength
calibration solutions. The transformations are used to convert the 2D image into a 3D data cube, consisting of the slice number, the position along the slice, and wavelength.
These data cubes, dubbed  ``ocubed'' files by the KDERP,  are used in the subsequent steps. 

The sky background in the science exposures is determined from the offset sky exposures. The sky frames cannot be used directly as the sky spectrum changes significantly over the 20-minute interval between successive exposures. Instead, we model the {\em variation} in the sky spectrum with a principal component analysis (PCA). The method is introduced and explained in detail in van Dokkum et al.\ (2019). Briefly, 1D sky spectra are extracted from the nine offset exposures by averaging over the two spatial dimensions after masking GC-39 and serendipitous objects in the field. These nine spectra are analyzed with singular value decomposition using the {\tt scikit-learn} package, with six components. These eigenspectra,  along with the average of the nine spectra and an approximate model for the galaxy spectrum, are then fitted to 1D extractions of each of the eight science exposures. 
The 1D sky model for each science exposure is  subtracted from each spatial pixel in the science data cube.  We note that
the model does not take possible spatial variation in the sky into account and is 
insensitive to offsets that are not correlated with specific sky emission or absorption
features (see \S\,\ref{sec:dispersion}).

\begin{figure*}[t!]
{\centering
  \includegraphics[width=180mm]{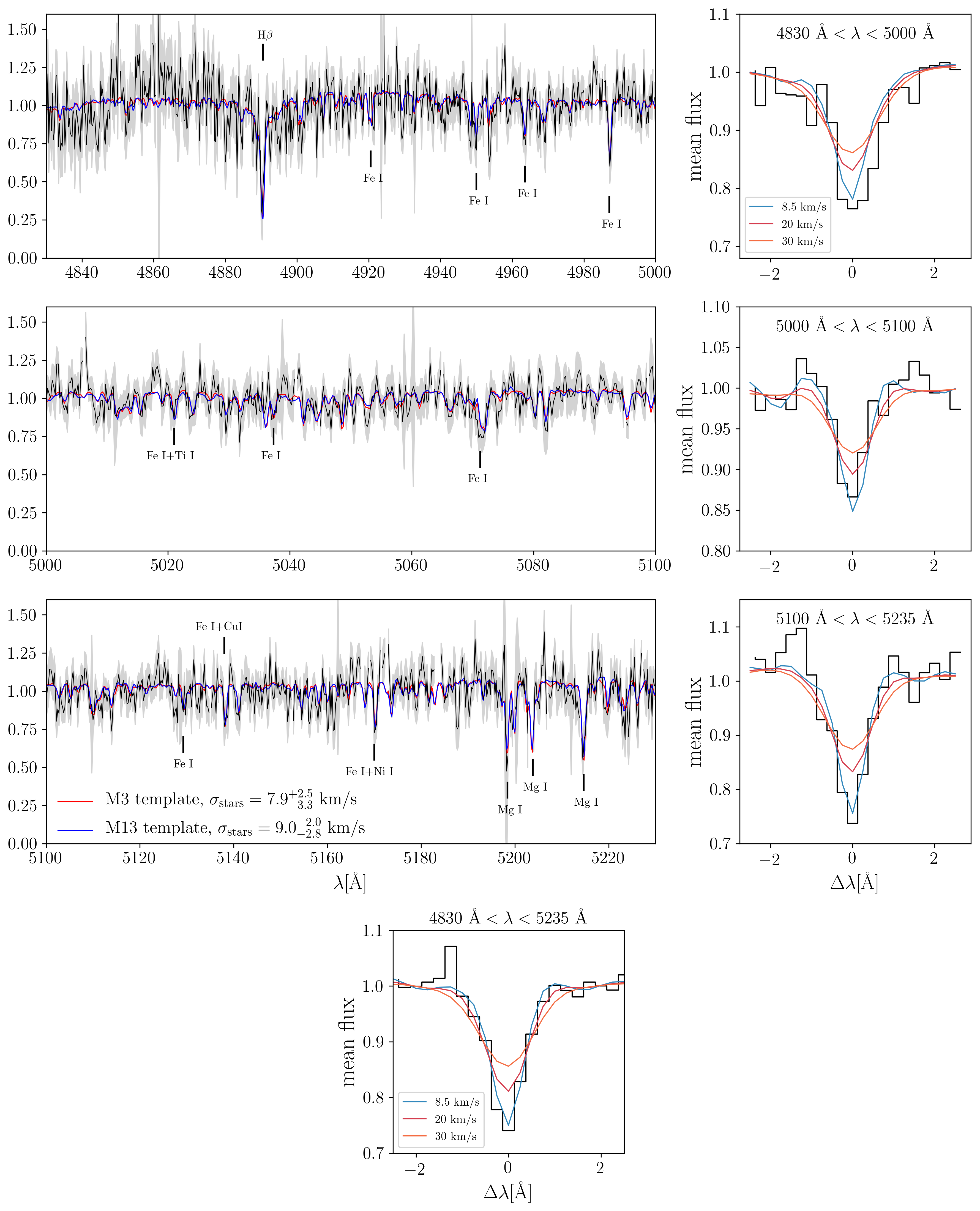}
  \caption{{\em Main panels:} Integrated 2.67 hrs KCWI spectrum of NGC1052-DF2 (black), with $1\sigma$ uncertainties (gray). The best-fits of the two empirical M3 and M13 templates (see Section \ref{sec:template}) that were used to determine the kinematics are shown in red and blue, respectively. The high resolution of KCWI allows us to detect a large number of absorption lines with a high accuracy. {\em Right:} Average of the 10 strongest absorption features in each spectral region, along with the best-fitting models and models with a higher dispersion. {\em Bottom:} Average of the 20 strongest lines in the entire spectrum. }
  \label{fig:spectrum}
}
\end{figure*}

Finally, 1D combined spectra for different spatial regions are created by extracting them from the individual science cubes and averaging them with optimal weighting. In the combination step pixels that deviate $>3\sigma$ from the median are not included in the average. Ten spatial regions are extracted: the sum over the entire field and nine
rectangular regions in a $3\times 3$ grid (see \S\,\ref{sec:dispersion}).
The summed spectrum is shown in Figure \ref{fig:spectrum}; selected redshifted absorption lines are marked.

\section{Kinematics} \label{sec:dispersion}

\subsection{Empirical Templates}
\label{sec:template}

The stellar kinematics of NGC1052-DF2 are measured by fitting template spectra to the extracted 1D spectrum. Key requirements are that the resolution of the template is well characterized and that template mismatch is minimized. This is not easily accomplished, given the high resolution of our data ($\approx 12$\,\kms at $\lambda=5000$\,\AA) and the low metallicity of the stellar population. We resolve these issues by using spatially-integrated spectra of old, metal poor Galactic globular clusters as templates, obtained with the same instrumental configuration.

The Galactic globular clusters M3 and M13 were observed on April 17, 2018. The metallicities of these clusters are ${\rm [Fe/H]}\approx -1.5$ (\citealt{1996AJ....112.1487H}), slightly lower than the expected metallicity of NGC1052-DF2 based on its stellar mass and slightly higher than that based on its velocity dispersion (\citealt{2013ApJ...779..102K}, \citealt{2018ApJ...859...37G}). The total integration time was 600\,s on each cluster, and 600\,s on a nearby sky field. The data reduction and spectral extraction followed the same procedures as described in \S\,\ref{sec:reduction} for NGC1052-DF2; this ensures that any instrument-induced effects (such as small spatial variations in the wavelength calibration) are in common between the templates and the science data. The spectra are simple averages of the entire KCWI field, with individual stars contributing at most a few percent. We verified that the instrumental resolution, as measured from arc lamps, is consistent between the April globular clusters data and the October NGC1052-DF2 data.

\begin{figure*}[t!]
{\centering
  \includegraphics[width=180mm]{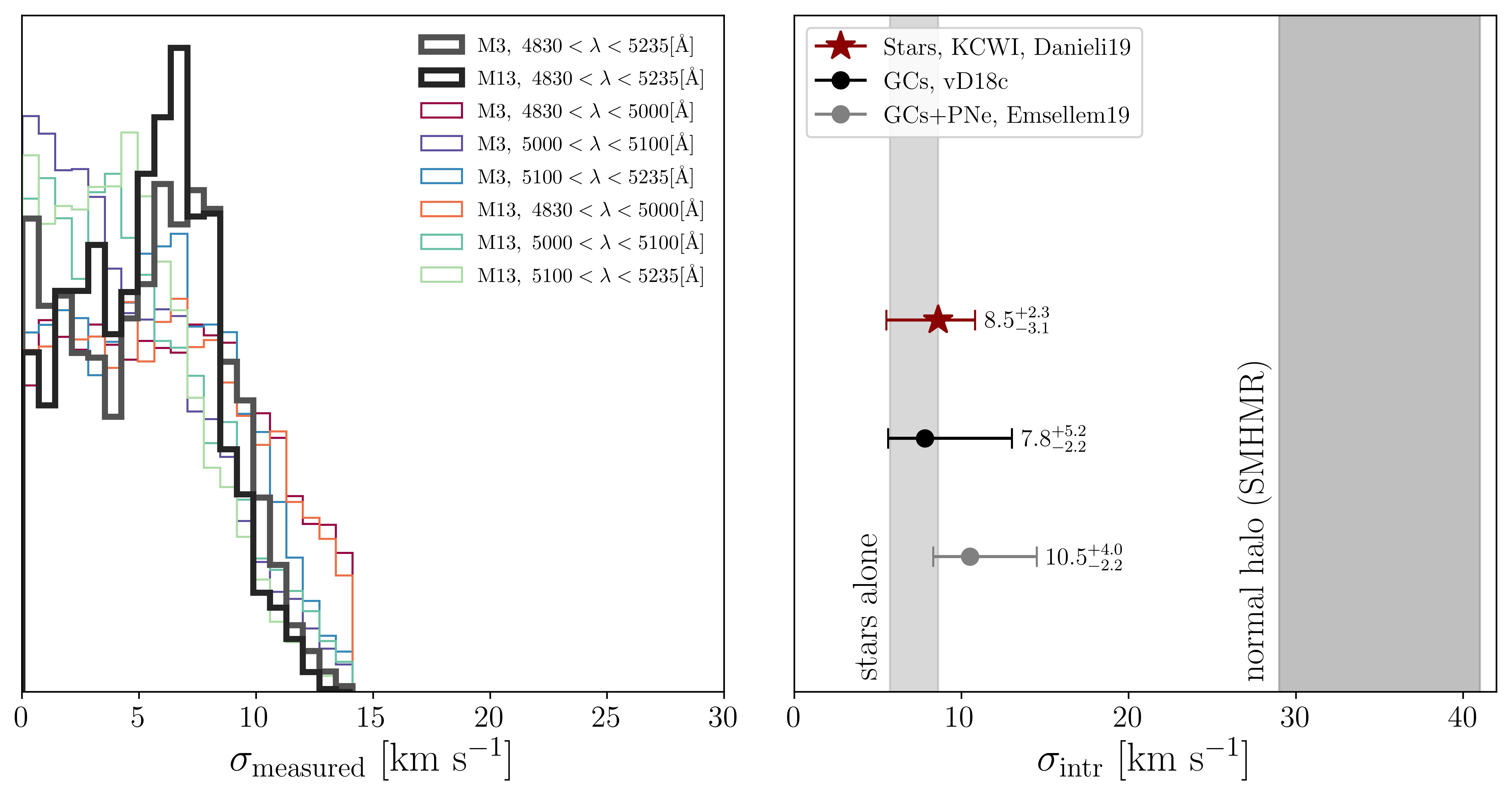}
  \caption{\textit{Left panel}: Posterior from eight MCMC runs of fitting the KCWI spectrum with two different templates and four different wavelength regimes.
The upper limit is very well constrained in all MCMC runs with the full wavelength range giving the strongest constraints, as expected. \textit{Right panel:} Constraints on the intrinsic velocity dispersion of NGC1052-DF2. The stellar velocity dispersion measured in this study (dark red star) is consistent with the results obtained from ten globular clusters in \citet{2018RNAAS...2b..54V}. The stars alone contribute $\sigma_{\mathrm{stars}}=7.0^{+1.6}_{-1.3}$ \kms (light gray band) and the expectation from the stellar mass -- halo mass relation (\citealt{2013ApJ...770...57B}), assuming a standard NFW halo (\citealt{2001MNRAS.321..155L}), is $\sigma_{\mathrm{SMHM}}=35 \pm 6$ \kms.}
  \label{fig:results}
}
\end{figure*}

\subsection{Velocity Dispersion Measurement} \label{sec:disp}
The velocity dispersion was determined in the wavelength region 4830\,\AA\,$<\lambda<$\,5235\,\AA, using the M3 and M13 templates. The fit was performed with an implementation of the {\tt emcee} Markov Chain Monte Carlo sampler (\citealt{2013PASP..125..306F}), with the redshift and velocity dispersion as free parameters \citep[see][]{2016ApJ...828L...6V}. Besides a multiplicative polynomial we fit for a third-order additive polynomial, to account for both sky subtraction errors and template mismatch. Varying the order of this polynomial does not change the results significantly. When fitting the full spectral range we also fit for any subtle wavelength calibration mismatch between the template and the data, parameterized as an second-order polynomial with respect to the central wavelength. Although we find polynomial coefficients that are slightly different from zero, the resulting dispersion does not change when they are forced to zero.
The best fitting models are shown in Figure \ref{fig:spectrum} by the red and blue lines. The errors describe the differences between the data and the models well; the reduced $\chi^2$ values are 1.05 for the M3 template and 1.07 for the M13 template.
The measured dispersion is the quadratic difference between the velocity dispersion of NGC1052-DF2 and that of the globular clusters. To obtain the stellar dispersion of NGC1052-DF2 we correct the measured values:
\begin{equation}
\sigma_{\mathrm{stars}}^2 = \sigma_{\mathrm{meas}}^2+\sigma_{\mathrm{M3/M13}}^2,
\end{equation}
with $\sigma_{\mathrm{M3}} = 5.5 \pm 0.3$ \kms \ and $\sigma_{\mathrm{M13}} = 7.1 \pm 0.4$ \kms (\citealt{1996AJ....112.1487H}).

We find a stellar line-of-sight velocity dispersion of NGC1052-DF2 of $\sigma_{\mathrm{stars}}=7.9^{+2.5}_{-3.4}$\,\kms \ when fitting the M3 template and $\sigma_{\mathrm{stars}}=9.0^{+2.0}_{-2.8}$\,\kms \ when fitting the M13 template. These numbers are in excellent agreement. The mean is $\sigma_{\mathrm{stars}}=8.5^{+2.3}_{-3.1}$ \kms. We note that the lower bound of 5.4 \kms \ is somewhat artificial, as it is partially determined by the internal dispersion of the globular clusters. The MC samples extend all the way to 0 \kms \ (see Figure \ref{fig:results}). The 95\,\% confidence upper limit on the dispersion is 11.8 \kms.
The central velocity dispersion for M13 is also somewhat uncertain; Kamann et al (\citeyear{2014A&A...566A..58K}) find a higher value than \citealt{1996AJ....112.1487H}, although they note that dispersions in this regime cannot be measured reliably given the instrumental resolution of PMAS ($\sigma_{\rm instr}\approx 18$\,\kms). An over- or underestimation of the intrinsic dispersion of M13 would result in a slightly different inferred dispersion for NGC1052-DF2. In that context, it is reassuring that the two independent measurements using the M3 and M13 templates are fully consistent with each other.

We performed two further tests of the stability of this result.
First, the spectrum was split into three wavelength regions. The first region is dominated by H$\beta$, the second by relatively weak Fe lines, and the third by Mg (see Figure \ref{fig:spectrum}). The MCMC posteriors for these fits are shown in the left panel of Figure \ref{fig:results}, and the corrected velocity dispersions are listed in Table \ref{table:dispersions}.
All inferred dispersions are consistent within 1.5\,\kms.
Next,  we split the data into nine spatial bins and fitted those independently. In all cases, the best-fit dispersion is well within $1\sigma$ of the value from fitting the full wavelength range.
An additional test we performed was fitting the M3 spectrum with the M13 spectrum as a template. The measured dispersion of M3 is consistent with zero and the intrinsic dispersion is consistent with the value from the literature (\citealt{1993ASPC...50..357P}, \citealt{1996AJ....112.1487H}). 

The robustness of our results is illustrated by the small panels in Figure \ref{fig:spectrum}, where we show the average observed absorption in the spectral regions corresponding to the strongest absorption lines in the templates. The 8.5\,\kms\ model is an excellent fit for all wavelength regions and also for the average of the 20 strongest lines in the entire spectrum. Measuring velocity dispersions in the 10\,\kms\,--\,30\,\kms\ regime is well-suited to KCWI.

The inferred intrinsic stellar velocity dispersion is consistent with the constraints on the velocity dispersion derived using globular clusters in \citet{2018RNAAS...2b..54V} with $\sigma_{\rm gc}=7.8^{+5.2}_{-2.2}$ \kms \ and in \citet{2018arXiv181207345E} with an estimated value of $\sigma_{\rm gc}=10.5^{+4.0}_{-2.2}$ \kms. 
These various results are shown in 
the right panel of Figure \ref{fig:results}.

\noindent
\begin{deluxetable}{cccccc} 
\tablecaption{NGC1052-DF2 Stellar Dispersions\label{table:dispersions}}
\tablehead{\colhead{Template} & \colhead{$\lambda$} & \colhead{$\sigma_{\mathrm{meas}}$} & \colhead{$\sigma_{\mathrm{stars}}$}
\\
  & [$\mathrm{\AA}$] & [km\,s$^{-1}$] & [km\,s$^{-1}$]}
\startdata
        M3 & $[4830,5235]$ & $5.8^{+3.1}_{-4.1}$ & \bf{$7.9^{+2.5}_{-3.4}$} \\
        M13 & $[4830,5235]$ & $5.5^{+2.5}_{-3.6}$ & \bf{$9.0^{+2.0}_{-2.8}$} \\
        M3 & $[4830,5000]$ & $6.4^{+4.7}_{-4.4}$ & $8.5^{+4.0}_{-3.7}$ \\ 
        M3 & $[5000,5100]$ & $4.1^{+4.1}_{-2.9}$ & $6.9^{+3.2}_{-2.3}$ \\
        M3 & $[5100,5235]$ & $5.7^{+3.7}_{-3.9}$ & $7.9^{+3.1}_{-3.2}$ \\
        M13 & $[4830,5000]$ & $6.2^{+4.7}_{-4.2}$ & $9.4^{+3.8}_{-3.4}$ \\ 
        M13 & $[5000,5100]$ & $4.4^{+4.2}_{-3.1}$ & $8.4^{+3.2}_{-2.3}$ \\
        M13 & $[5100,5235]$ & $4.3^{+3.3}_{-2.9}$ & $8.3^{+2.5}_{-2.2}$ 
\enddata
\end{deluxetable}

\subsection{Systemic Velocity and Stellar Velocity Field}

\begin{figure*}[t!]
{\centering
  \includegraphics[width=180mm]{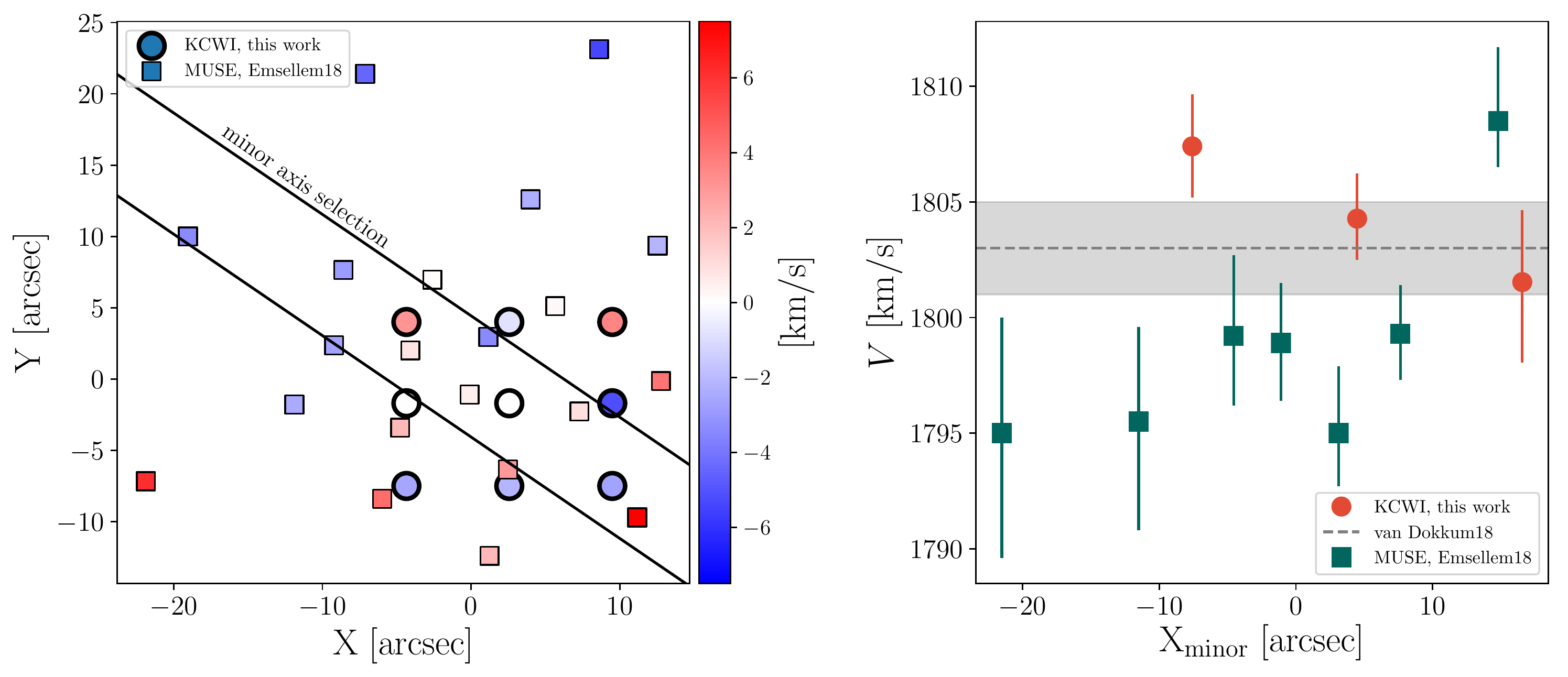}
  \caption{\textit{Left panel:} velocity field measured from stars in different spatial positions of the galaxy, as measured by MUSE (squares; \citealt{2018arXiv181207345E}) and KCWI (circles). The velocities are relative to the mean velocity in each dataset. \textit{Right panel:} absolute velocities in a band (black diagonal lines) roughly corresponding to the minor axis. There is an offset between the MUSE and KCWI velocities, and no clear gradient in our data.}
  \label{fig:velocity}
}
\end{figure*}

Besides the velocity dispersion we also obtain  
a measurement of the mean systemic velocity. The best-fit values for the M3 and M13 templates are $v_{\mathrm{stars}} = 1805.2^{+1.1}_{-1.1}$ \kms and $v_{\mathrm{stars}} = 1804.7^{+1.0}_{-1.1}$ \kms, respectively. The two values are consistent with each other and also with the mean velocity of the ten globular clusters as measured in \citet{2018Natur.555..629V}: $\langle v_{\rm gc} \rangle=1803^{+2}_{-2} $ \kms.

Next, we examine the systemic velocities in the nine spatial bins described in \S\,\ref{sec:disp}.
The rms scatter among the nine velocities is 2.8\,\kms. This is very similar to the mean velocity uncertainty (2.2\,\kms), leaving little room for velocity gradients of the same order as the velocity dispersion.
In the left panel of Figure \ref{fig:velocity} we show the stellar velocity field.
We find no clear gradient in the velocities measured from our data within our FOV. This is in contrast to results from 
\citet{2018arXiv181207345E}, who report a gradient of $2.8 \pm 0.9$ \kms \ per 10'' along the minor axis. In the right panel of Figure \ref{fig:velocity} we compare the absolute velocities of several spatial bins along the minor axis directly, as were obtained from our data (orange circles) and from the MUSE data (green squares). We shift the MUSE velocities by $5.4 $ \kms \ to account for the different applied redshift-velocity transformation ($cz$ in our analysis versus $c\ln (1+z)$ in \citealt{2018arXiv181207345E}).
If there is a slight trend in the KCWI data, it is in the opposite direction from that seen in the MUSE data. Given the large offset in the absolute velocities between the MUSE results on the one hand and the KCWI, LRIS, and DEIMOS results on the other, and the lack of a clear trend in our data, we do not confirm the presence of the gradient claimed in \citet{2018arXiv181207345E}. \footnote{These authors have also measured the stellar velocity dispersion, but  owing to the complexities of measuring well below the instrumental  resolution, the final values are not yet known at the time of writing and we therefore cannot compare them directly to ours (E. Emsellem, private communication).}
We note that the MUSE velocities are consistent with ours in the lower (Southern) part of the MUSE data cube. 




\section{Discussion} \label{sec:discussion}

\begin{figure*}[t!]
{\centering
  \includegraphics[width=\textwidth]{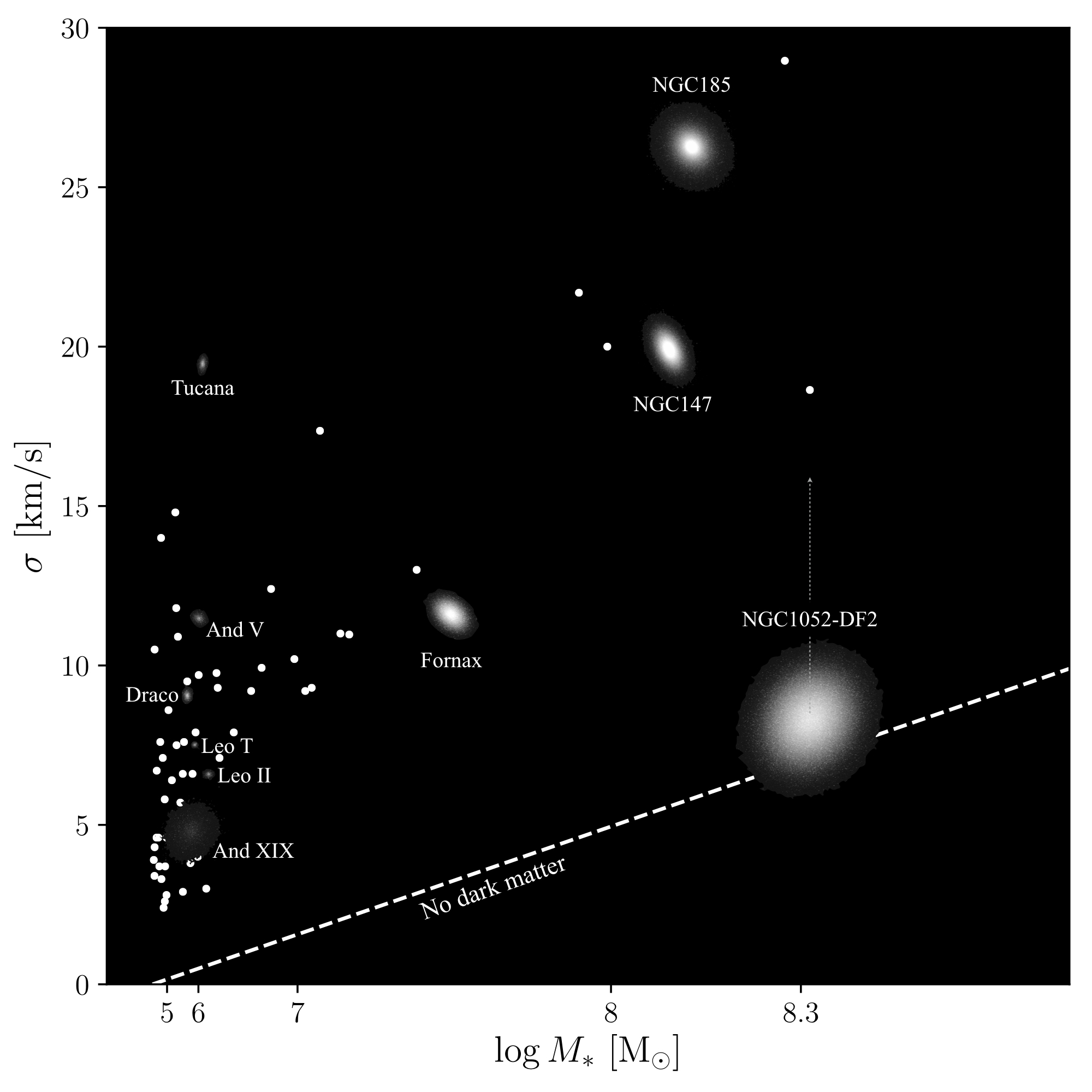}
  \caption{Local Group dwarf galaxies (white dots) and NGC1052-DF2 in the velocity dispersion--stellar mass plane. NGC1052-DF2 and several illustrative Local Group galaxies were modeled and placed at $20 \ \mathrm{Mpc}$ (using the ArtPop code, \citealt{2018ApJ...856...69D}). Its large size and relatively high stellar mass, along with its low velocity dispersion, distinguish it from Local Group galaxies. The dashed line is the approximate relation between stellar mass and velocity dispersion in the absence of dark matter.}
  \label{fig:mass}
}
\end{figure*}

In this \textit{Letter} we have presented stellar kinematics measurements for the galaxy NGC1052-DF2, using high resolution ($\sim 12 $ \kms) integral-field spectroscopy with the Keck Cosmic Web Imager (KCWI) on Keck II. 
We measure a systemic velocity of $\langle v_{\mathrm{stars}} \rangle=1804.9^{+1.0}_{-1.1}$\,\kms, and confirm that the ten star clusters that were previously used to constrain the kinematics of NGC1052-DF2 (\citealt{2018RNAAS...2b..54V}, \citealt{2018ApJ...863L..15W}) are indeed associated with the diffuse stellar light of the galaxy.
Thanks to the exceptionally high resolution of KCWI, we are providing a robust measurement of the stellar velocity dispersion of NGC1052-DF2.
We measure  $\sigma_{\mathrm{stars}}=8.5^{+2.3}_{-3.1}$  \kms \ within the effective radius, consistent with the revised dispersion of $\sigma_{\mathrm{gc}}=7.8^{+5.2}_{-2.2}$ \kms \ measured from the ten globular clusters using the maximum likelihood method in \citet{2018RNAAS...2b..54V}. 

In Figure \ref{fig:mass} we show the distribution of Local Group galaxies in the plane of velocity dispersion versus stellar mass, using the compilation of \citet{2012AJ....144....4M}. Several galaxies are displayed as model images, created with the ArtPop code (\citealt{2018ApJ...856...69D}). This Figure graphically illustrates the unusual nature of NGC1052-DF2: the galaxy combines a relatively high stellar mass with a large size and a very low velocity dispersion. The dashed line is an indicative relation between stellar mass and velocity dispersion in the absence of dark matter, for the radial regime where the dispersion profile is approximately isothermal: $\sigma \sim 5\times10^{-4} (M_*/{\rm M}_{\odot})^{0.5}$ \kms. Typical dwarf galaxies fall above the line, as they are dark matter dominated, but NGC1052-DF2 is on the line within the errors. 

We quantify this by using the newly measured stellar velocity dispersion along with the projected circularized half-light radius of $R_{\rm e,c}=2.0$\,kpc\ (\citealt{2018ApJ...868...96C}) to determine the dynamical mass of NGC1052-DF2 within the 3D half-light radius $r_{1/2}\approx 4/3 R_{\rm e,c}$ (\citealt{2010MNRAS.406.1220W}). We find $M(r<r_{1/2})=1.3 \pm 0.8 \times 10^8 \ \mathrm{M}_{\odot}$. The stellar mass within the half light radius is $M_{\rm stars}(r<r_{1/2})=1.0^{+0.2}_{-0.2} \times 10^8$\,M$_{\odot}$
(see \citealt{2018Natur.555..629V}), and we infer that the dynamical mass is consistent with the mass in stars alone.
We refer to \citet{2018ApJ...863L..15W} for quantitative constraints on the halo mass as derived from the globular clusters; our measurement confirms the central assumption in \citet{2018ApJ...863L..15W} that the globular clusters indeed trace the potential of NGC1052-DF2.
We note that if NGC1052-DF2 is a thin rotating disk seen close to face-on, its axis ratio of 0.85 implies inclination-corrected velocities that are (at most) a factor of 1.9 higher than the observed ones (\citealt{2018Natur.555..629V}). This scenario is unlikely given the lack of detected rotation along the major axis or in the globular clusters, and the discovery of a second galaxy missing dark matter, NGC1052-DF4, in the same group (\citealt{2019arXiv190105973V}). Nevertheless, for consistency with the other data points, we show this inclination correction with a dotted line in Figure \ref{fig:mass}.

Our study confirms that NGC1052-DF2 has far less dark matter than expected, and perhaps no dark matter at all. 
Future studies can examine what physical processes and formation schemes can result in this deficiency of dark matter on kpc scales. This is particularly challenging given that other similar-looking UDGs appear to have normal (or even ``overmassive'') halos (see, e.g., \citealt{2018ApJ...856L..31T}).
It is now critical to determine whether NGC1052-DF2 is a unique galaxy or whether this ``missing dark matter problem'' is relatively common. If it is, it implies that the scatter in stellar mass at low halo masses is extremely large (see, e.g., \citealt{2018arXiv180607893B}). 




Another essential question is whether other properties of NGC1052-DF2, such as its nature as a UDG and its intriguing population of globular clusters, are related to its dark matter deficiency.
Finding a closer-by system with a low velocity dispersion would allow us to constrain its properties (even) more accurately, and place
strong constraints on dark matter and halo models.

\acknowledgments
We thank Luca Rizzi for his help with the KCWI pipeline.
Support from STScI grants HST-GO-13682 and HST-GO-14644, as well as NSF grants AST-1312376, AST-1613582 and NSF AST-1616170, is gratefully acknowledged. AJR is supported as a Research Corporation for Science Advancement Cottrell Scholar, and via a NASA
Keck PI Data Award.


\end{document}